  [2004/08/25 v0 HiX 2004, Marseille - writeup-version driver (PGR)]
\documentclass[
    ,final            
    ,sort&compress    
    ,british          
  ]
  {aipproc}
\layoutstyle{6x9}
\usepackage{Ratcliffe}
\title{Semi-Inclusive DIS and Transversity}
\author{Philip G. Ratcliffe}{%
  address={%
    Dipartimento di Fisica e Matematica,
    Universit\`{a} degli Studi dell'Insubria \\
    and \\
    Istituto Nazionale di Fisica Nucleare,
    Sezione di Milano
  }
}
\begin{fmffile}{Ratcliffe-fmf}
\begin{document}

\begin{abstract}
A review is presented of some aspects of semi-inclusive \acl{DIS} and
transversity. In particular, the role of $k_T$-dependent and higher-twist (or
multi-parton) distributions in generating single-spin asymmetries is discussed.
\end{abstract}

\maketitle

\section{Introduction}

A large part of the material presented here is gleaned from
\cite{Barone:2001sp, Barone:2003fy}, where the coverage is more complete.
Therefore, much credit and thanks go to my two collaborators: Enzo Barone and
Alessandro Drago. Further (less condensed and more complete) conference reviews
may also be found in other proceedings \cite{Ratcliffe:2002qb,
Artru:2002pu}.
Finally, the space available forces a limited choice of topics and there are,
unfortunately, many I cannot even mention.

By way of motivation for the interest, let me recall that transversity is the
last piece in the partonic jig-saw puzzle of the hadron. A number of
experiments aim at its study: HERMES, COMPASS and the RHIC spin programme. And
while the \ac{QCD} theory describing transversity is now solid, transverse-spin
effects are notoriously surprising; consider, \eg, the unexpectedly large
\acp{SSA}.


Transversity then is the third and final leading-twist ($\tau=2$) partonic
distribution function. Now, it is important to distinguish between partonic
distributions: $q(x)$, $\DL{q}(x)$, $\DT{q}(x)$ \etc and \ac{DIS} structure
functions: $F_1$, $F_2$, $g_1$, $g_2$ \etc. In both the unpolarised and
helicity-dependent cases at leading twist there is a simple correspondence
between the two: \ac{DIS} structure functions are little more than weighted
sums of partonic distributions (or densities). However, in the transverse-spin
case, firstly, there is no \ac{DIS} transversity structure function and,
secondly, $g_2$ does not correspond to a partonic density.

The parton-model description provides a simple probabilistic view of hadron
structure (herein I shall use $f_1$, $g_1$ and $h_1$ generically):
\begin{description}
\item
[$f_1(x)$ or $q(x)$] represent the probability of finding a given parton type
with light-cone momentum fraction $x$ inside a given parent hadron;
\item
[$g_1(x)$ or $\DL{q}(x)$] the same but weighted by parton helicity relative to
parent helicity;
\item
[$h_1(x)$ or $\DT{q}(x)$] weighted by transverse-spin projection relative to
parent transverse-spin direction.
\end{description}
Note, however, $h_1(x)$ does not measure quark transverse polarisation; $g_2$
has this role.

Turning now to \acp{SSA}, they generically reflect correlations of the form
$\Vec{s}\cdot(\Vec{p}\vprod\Vec{k})$, where $\Vec{s}$ is a particle spin
vector, $\Vec{p}$ and $\Vec{k}$ are initial/final particle/jet momenta; for
example, $\Vec{s}$ might be a target polarisation vector (transverse),
$\Vec{p}$ a beam direction and $\Vec{k}$ a final-state particle direction.
Thus, spins involved in \acp{SSA} are typically transverse (however, there are
exceptions). Transforming the spin basis from transversity to helicity such an
asymmetry takes on the schematic form, using
$\ket{\uparrow/\downarrow}=\frac1{\sqrtno2}[\ket{+}\pm\I\ket{-}]$,
\begin{align}
  \mathcal{A}_N
  \sim
  \frac{\braket{\uparrow|\uparrow}-\braket{\downarrow|\downarrow}}
       {\braket{\uparrow|\uparrow}+\braket{\downarrow|\downarrow}}
  \sim
  \frac{2\Im\braket{+|-}}{\braket{+|+}+\braket{-|-}}.
\end{align}
The appearance of both $\ket{+}$ and $\ket{-}$ in the numerator indicates a
spin-flip amplitude. Indeed, interference between a spin-flip and a non-flip
amplitude, with a relative phase difference, is necessary. It was realised
early on \cite{Kane:1978nd} that in the Born approximation and massless (or
high-energy) limit a gauge theory, such as \ac{QCD}, cannot furnish either
requirement: fermion helicity is conserved and tree diagrams are real. This
naturally led to the claim \cite{Kane:1978nd} that ``\dots\ \emph{observation of
significant polarizations in the above reactions would contradict either
\acs{QCD} or its applicability.}''

Now, although large experimental asymmetries were found, \ac{QCD} survived! A
way out was discovered \cite{Efremov:1985ip} by considering the three-parton
correlators involved in $g_2$: the relevant mass scale when considering
helicity flip is not the quark mass but a hadronic mass; and the pseudo
two-loop nature of the diagrams leads to an imaginary part in certain regions
of partonic phase space. However, it took some time before progress was made
and the richness of the available structure was fully exploited (see,
\eg,~\cite{Qiu:1991pp.x}).

\section{A Brief History of Transversity}

\begin{wraptable}{r}{0pt}
  \makeatletter
    \let\tablehead\AIP@tablehead
    \let\@makecaption\AIP@maketablecaption
    \AIPtablefont
  \makeatother
  \quad
  \begin{minipage}{55mm}
  \centering
  \vspace*{-0.5ex}
  \caption{%
    The twist classification (up to $\tau=4$) for the various parton
    distributions, including spin dependence (both transverse and
    longitudinal).
  }
  \label{tab:twist-class}
  \vspace*{1.0ex}
  \(
  \begin{array}{l|ccc}
    \multicolumn{1}{r|}{\hspace*{5em}$twist$} & 2 & 3 & 4
  \\
  \hline \vphantom{\displaystyle\frac12}
    $unpolarised$  & f_1 &  e  & f_4
  \\
    $longitudinal$ & g_1 & h_L & g_3
  \\[0.6667ex]
    $transverse$   & h_1 & g_T & h_3
  \end{array}
  \)
  \end{minipage}
\end{wraptable}
Quark \emph{transversity} (the concept though not the term\footnote{For very
early use of the term \emph{transversity}, see \cite{Kotanski:1966a1.x}.}) was
introduced by \citet*{Ralston:1979ys} in the \ac{DY} process. An important
clarification of transversity, the role of chirally-odd parton distributions
and the general twist classification was provided by \citeauthor*{Jaffe:1992ra}
in \cite{Jaffe:1992ra}\footnote{The term \emph{transversity}, following
\cite{Goldstein:1989jy}, was also suggested to distinguish it from transverse
spin.}, see Table~\ref{tab:twist-class}. As twist runs from 2 to 4, the number
of ``bad'' light-cone components runs from 0 to~2.
\\ \indent
The \ac[\hyphenate]{LO} anomalous dimensions for transversity were first
calculated very early on in \cite{Baldracchini:1981uq}, which went unnoticed,
and later re-calculated in \cite{Artru:1990zv}. They were also calculated (as
part of the $g_2$ evolution) in \cite{Kodaira:1979ib, Antoniadis:1981dg,
Bukhvostov:1983te, Ratcliffe:1986mp}. The \ac[\hyphenate]{NLO} \ac{DY}
coefficient functions were calculated in \cite{Vogelsang:1993jn,
Contogouris:1994ws} while the \ac{NLO} anomalous dimensions were calculated in
\cite{Hayashigaki:1997dn, Kumano:1997qp, Vogelsang:1998ak}. The effects of
evolution have been studied by many authors---for more details and a general
review see, \eg,~\cite{Barone:2001sp}.

\section{Transverse-Spin Basics}

\begin{wrapfigure}{r}{0pt}
  \makeatletter
    \let\@makecaption\AIP@makefigurecaption
  \makeatother
  \quad
  \begin{minipage}[t]{42mm}
  \vspace*{-0.5ex}
  \centering
  \includegraphics[width=38mm,bb=336 584 486 679,clip]{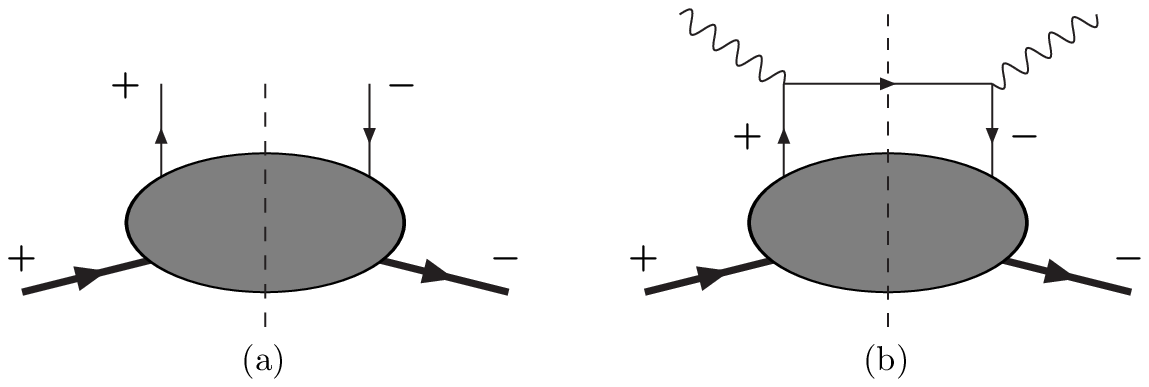}
  \caption{%
    Chirality-flip \ac{DIS} diagram forbidden by helicity conservation.
  }%
  \label{fig:DIS-forbid}
  \vspace*{-3.2cm}
  \end{minipage}
\end{wrapfigure}
Transversity is one of three twist-two structures:
\begin{align}
  q(x)
  &=
  \int \! \frac{\D\xi^-\mkern-7mu}{4\pi} \, \E^{\I xP^+ \xi^-}
  \braket{ PS |
    \anti\psi(0)
      \gamma^+
    \psi(0, \xi^-, \Vec{0}_\perp)
  | PS },
  \nonumber
\\[-1ex]
  \DL{q}(x)
  &=
  \int \! \frac{\D\xi^-\mkern-7mu}{4\pi} \, \E^{\I xP^+ \xi^-}
  \braket{ PS |
    \anti\psi(0)
      \gamma^+\gamma_5
    \psi(0, \xi^-, \Vec{0}_\perp)
  | PS },
  \nonumber
\\[-1ex]
  \DT{q}(x)
  &=
  \int \! \frac{\D\xi^-\mkern-7mu}{4\pi} \, \E^{\I xP^+ \xi^-}
  \braket{ PS |
    \anti\psi(0)
      \gamma^+\gamma^1\gamma_5
    \psi(0, \xi^-, \Vec{0}_\perp)
  | PS }.
\end{align}
Here the $\gamma_5$ matrix signals spin dependence while the extra $\gamma^1$
matrix in $\DT{q}(x)$ signals chirality flip; this last precludes transversity
contributions in \ac{DIS}, see Fig.~\ref{fig:DIS-forbid}.

\Ac{QCD} and electroweak vertices conserve quark chirality. Thus,
charged-current inter-
\begin{wrapfigure}{l}{0pt}
  \makeatletter
    \let\@makecaption\AIP@makefigurecaption
  \makeatother
  \begin{minipage}{7cm}
  \centering
    \begin{fmfgraph*}(20,20)
      \fmfbottom{i1,o2}
      \fmftop{o1,i2}
      \fmf{fermion,label=$+$,l.side=left}{i1,v1}
      \fmf{fermion,label=$+$,l.side=left}{v1,o1}
      \fmf{fermion,label=$-$,l.side=left}{i2,v2}
      \fmf{fermion,label=$-$,l.side=left}{v2,o2}
      \fmffreeze
      \fmf{gluon}{v1,v2}
      \fmfiv{label=(a)}{(0.5w,0.25h)}
    \end{fmfgraph*}%
  \hspace{3.5em}
    \begin{fmfgraph*}(20,20)
      \fmfbottom{i1,o2}
      \fmftop{o1,i2}
      \fmf{fermion}{i1,v1}
      \fmf{gluon}{v1,o1}
      \fmf{gluon}{i2,v2}
      \fmf{fermion}{v2,o2}
      \fmffreeze
      \fmf{fermion,label=$+$,l.side=left}{i1,v1}
      \fmf{fermion,label=$-$,l.side=left}{v2,o2}
      \fmf{fermion,label=?,l.side=left}{v1,v2}
      \fmfiv{label=(b)}{(0.5w,0.25h)}
    \end{fmfgraph*}
  \caption{%
    (a) Evolution kernel for transversity.
    (b) Disallowed gluon--fermion mixing kernel.
  }%
  \label{fig:evolution}
  \end{minipage}
  \hspace*{1em}
\end{wrapfigure}
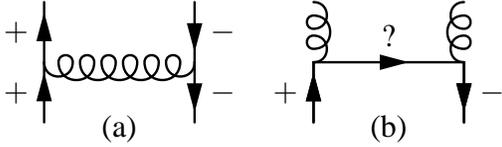
actions exclude transversity, since only a single chirality interacts. Note
though that chirality flip is not a problem if the quarks connect to different
hadrons (as in \ac{DY}). However, a caveat to measuring transversity in \ac{DY}
is that the azimuth of the lepton pair be left unintegrated. The same
observations lead to another important consequence: the \ac{LO} \ac{QCD}
evolution of transversity is of the non-singlet type, see
Fig.~\ref{fig:evolution}.

\section{Transversity: Models and \acs{QCD}}

Having opposite charge-conjugation properties, $\DL{q}$ and $\DT{q}$ are not
simply related. Decomposing $\DL{q}$ as $\DL{q}^\text{NS}+\DL{q}^\text{S}$, one
might imagine $\DT{q}\simeq\DL{q}^\text{NS}$. In the non-relativistic quark
model this is the case. However, in a relativistic model the lower quark
wave-function components spoil the identity and, \eg, the MIT bag gives
\cite{Jaffe:1992ra}:
\begin{align}
  \DL{q}^\text{NS}
  = \textstyle
  c \int r^2\D{r} \, (f^2 - \tfrac13 g^2)
  \quad\text{while}\quad
  \DT{q}
  = \textstyle
  c \int r^2\D{r} \, (f^2 + \tfrac13 g^2),
\end{align}
where $f$, $g$ (the upper, lower components) contribute differently due to the
extra $\gamma_0$.

By considering hadron--parton amplitudes \citet{Soffer:1995ww} derived an
intriguing bound: $|\DT{q}(x)|\le{q}_+(x)$ or
$2|\DT{q}(x)|\le{q}(x)+\DL{q}(x)$. In \ac{QCD} the question arises as to the
effects of evolution on this bound \cite{Goldstein:1995ek}: its maintenance has
been checked explicitly to \ac{LO} in \cite{Barone:1997fh}, to \ac{NLO} in
\cite{Martin:1998rz} and discussed on more general grounds in
\cite{Bourrely:1998bx}; experimental verification thus becomes an important
test.

The \ac{LO} (non-singlet) DGLAP quark--quark splitting functions are
\begin{subequations}
\begin{align}
  \DL{P}_{qq}^{(0)}(x)
  &=
  P_{qq}^{(0)}(x)
  =
  \CF \left[ \frac{1+x^2}{1-x\,} \right]_+
  \qquad(\text{due to helicity conservation}),
\\
  \DT{P}_{qq}^{(0)}(x)
  &=
  P_{qq}^{(0)}(x) - \CF (1-x).
\end{align}
\end{subequations}
Thus, for both $P_{qq}^{(0)}$ and $\DL{P}_{qq}^{(0)}$ the first moments vanish
(leading to conservation laws and sum rules). The same is not true for
$\DT{P}_{qq}$ and so, there are no transversity sum rules.

\newcommand\taga{(a)}
\newcommand\tagb{(b)}
\begin{wrapfigure}{r}{0pt}
  \makeatletter
    \let\@makecaption\AIP@makefigurecaption
  \makeatother
  \quad
  \begin{minipage}{7cm}
  \centering
  \psfrag{0.5}{\mbox{}}
  \psfrag{1.5}{\mbox{}}
  \psfrag{2.5}{\mbox{}}
  \psfrag{3.5}{\mbox{}}
  \psfrag{(a)}{\tiny\taga}
  \psfrag{(b)}{\tiny\tagb}
  \psfrag{x}{$x$}
  \psfrag{u}{\tiny$u$}
  \psfrag{D}{\tiny$\Delta$}
  \psfrag{T}{\tiny$_T$}
  \psfrag{LO input}{\tiny LO input}
  \psfrag{NLO input}{\tiny\quad NLO input}
  \psfrag{LO evolution}{\tiny LO evolution}
  \psfrag{NLO evolution}{\tiny NLO evolution}
  \includegraphics[width=\textwidth,clip]{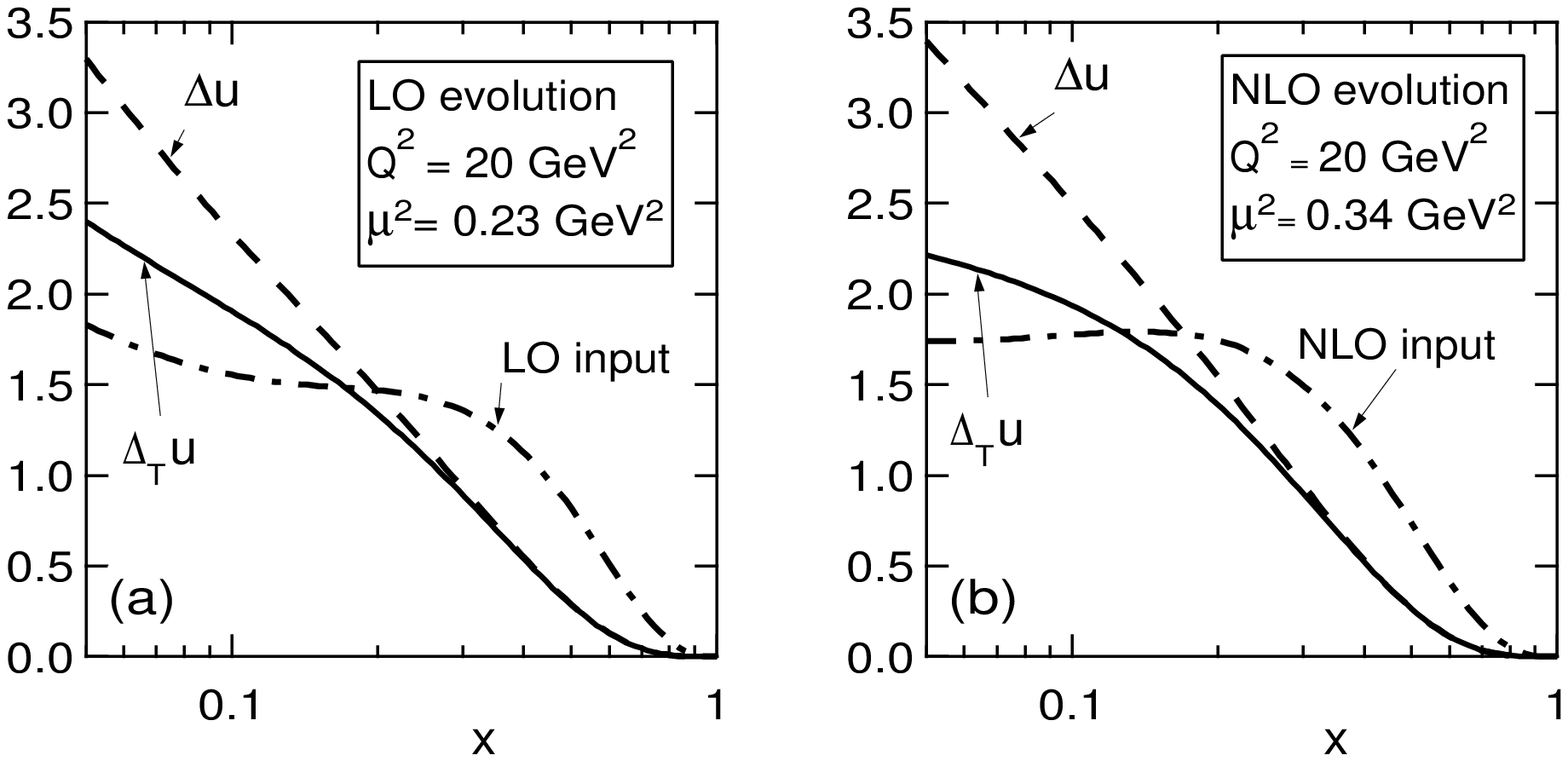}
  \caption{%
    The $Q^2$-evolution of $\DT{u}(x,Q^2)$ and $\DL{u}(x,Q^2)$ at \taga\
    \acs{LO} and \tagb\ \acs{NLO}, from~\cite{Hayashigaki:1997dn}.
  }
  \end{minipage}
\end{wrapfigure}
The importance of this difference has been studied both at \ac{LO} and
\ac{NLO}. The sign indicates that quark transverse polarisation decreases, as
compared to longitudinal polarisation. This is both bad news, since
transversity effects therefore die at very high energies (though only very
slowly), and good news, since evolution effects are stronger and therefore more
measurable (\ie, they are good for testing \ac{QCD}).

The chirally-odd nature of transversity means that at least two hadrons are
needed to probe transversity: $p^{\uparrow}p^\uparrow\to\mu^+\mu^-X$
\citep*{Ralston:1979ys, Cortes:1992ja}; $ep^\uparrow\to{e'\pi}X$
\citep{Collins:1993kk, Ji:1994vw, Gamberg:2003hf};
$pp^\uparrow\to\Lambda^{\uparrow}X$ \citep{deFlorian:1998am};
$ep^\uparrow\to\Lambda^{\uparrow}X$ \citep{Anselmino:2001js};
$ep^\uparrow\to{e'\pi^+\pi^-X}$ \citep{Ji:1994vw, Collins:1994kq, Jaffe:1998pv}
\etc. There are thus two basic categories: double- and single-spin asymmetries.
I shall now examine the latter more closely.

\section{Single-Spin Asymmetries}

\Acp{SSA} can be generated in various ways: higher-twist, $k_T$-dependent
distribution or fragmentation functions; or interference and vector-meson
fragmentation functions. Consider single-hadron production with a transversely
polarised beam or target:
\begin{align}
  A^\uparrow(P_A) \, + \, B(P_B) \, \to \, h(P_h) \, + \, X,
\end{align}
where $A$ is transversely polarised and the unpolarised (or spinless) hadron
$h$ is produced at large transverse momentum; \ac{PQCD} is thus applicable. One
measures the \ac{SSA}:
\begin{align}
  \mathcal{A}_{T}^h =
  \frac{\D\sigma(\Vec{S}_T) - \D\sigma(-\Vec{S}_T)}
       {\D\sigma(\Vec{S}_T) + \D\sigma(-\Vec{S}_T)}.
\end{align}
\begin{figure}[hbt]
  \centering
  \includegraphics[width=0.4\textwidth]{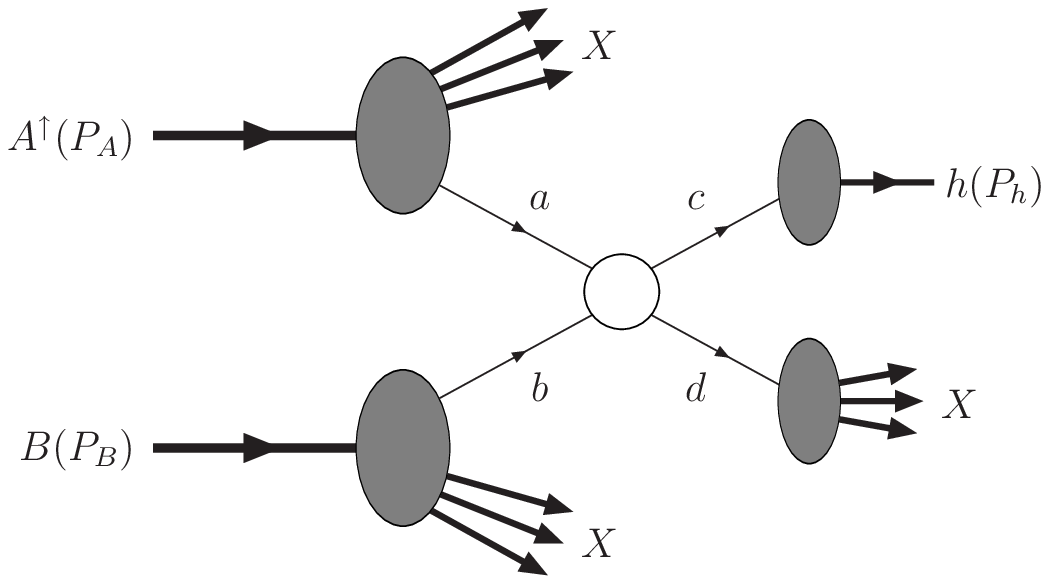}
  \qquad
  \includegraphics[width=0.3\textwidth]{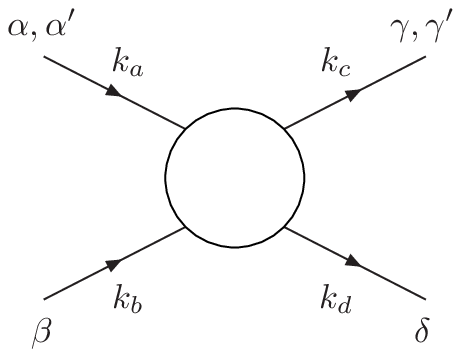}
  \caption{%
    Left: a
    schematic representation of single-hadron production with a transversely
    polarised beam or target.
    Right:
    the hard partonic scattering amplitude
    $\mathcal{M}_{\alpha\beta\gamma\delta}$ ($\alpha\dots\delta$ are
    Dirac indices).
  }
\end{figure}

According to the factorisation theorem, the differential cross-section for
semi-inclusive production may be written formally as
\begin{align}
  \qquad
  \D\sigma =
  \sum_{abc} \sum_{\alpha\alpha'\gamma\gamma'}
  \rho^a_{\alpha'\alpha} \,
  f_a(x_a) \otimes
  f_b(x_b) \otimes
  \D\hat\sigma_{\alpha\alpha'\gamma\gamma'} \otimes
  \mathcal{D}_{h/c}^{\gamma'\gamma}(z)
  ,
\end{align}
where $f_a$ ($f_b$) is the density of parton type $a$ ($b$) in hadron $A$
($B$), $\rho^a_{\alpha\alpha'}$ is the spin density matrix of parton $a$, and
$\mathcal{D}_{h/c}^{\gamma\gamma'}$ is the fragmentation matrix for parton $c$
into final hadron $h$. The elementary cross-section is
($\mathcal{M}_{\alpha\beta\gamma\delta}$ is the hard partonic scattering
amplitude)
\begin{align}
  \left(
    \frac{\D\hat\sigma}{\D\hat{t}}
  \right)_{\!\!\alpha\alpha'\gamma\gamma'}
  =
  \frac12 \, \sum_{\beta}
  \left(
    \frac{\D\hat\sigma}{\D\hat{t}}
  \right)_{\alpha\alpha'\beta\beta'\gamma\gamma'}
  =
  \frac1{16\pi\hat{s}^2} \, \frac12 \, \sum_{\beta\delta}
  \mathcal{M}^{\vphantom*}_{\alpha\beta\gamma\delta} \,
  \mathcal{M}^*_{\alpha'\beta\gamma'\delta}.
\end{align}
The off-diagonal elements of $\mathcal{D}_{h/c}^{\gamma\gamma'}$ vanish, \ie,
$\mathcal{D}_{h/c}^{\gamma\gamma'}\propto\delta_{\gamma\gamma'}$ when the
hadron produced is unpolarised. Helicity conservation then implies
$\alpha=\alpha'$, thus precluding any dependence on the polarisation of hadron
$A$. Either intrinsic quark transverse motion, or higher-twist effects can
circumvent this conclusion.

Quark intrinsic transverse motion can generate \acp{SSA} in three different
ways (always necessarily as $T$-odd effects):
\begin{enumerate}
\item
$\Vec\kappa_T$ in the final hadron $h$ allows
$\mathcal{D}_{h/c}^{\gamma\gamma'}$ to be non-diagonal (fragmentation level).
\item
$\Vec{k}_T$ in hadron $A$ requires $f_a(x_a)$ be replaced by
$\mathcal{P}_a(x_a,\Vec{k}_T)$, which may depend on the spin of $A$
(distribution level).$\vphantom{D_{/}\gamma}$
\item
$\Vec{k}'_T$ in hadron $B$ requires $f_b(x_b)$ be replaced by
$\mathcal{P}_b(x_b,\Vec{k}'_T)$. The transverse spin of $b$ in the unpolarised
$B$ may then couple to the transverse spin of $a$ (distribution level).
\end{enumerate}
The three mechanisms are: 1.~\citeauthor{Collins:1993kk} \cite{Collins:1993kk};
2.~\citeauthor{Sivers:1990cc} \cite{Sivers:1990cc}; and 3.~an effect studied in
by \citeauthor{Boer:1999mm} in \cite{Boer:1999mm}. All such
intrinsic-$\Vec\kappa_T$, -$\Vec{k}_T$, or -$\Vec{k}'_T$ effects are $T$-odd;
they require initial- or final-state interactions. Note that when quark
transverse motion is included, the \ac{QCD} factorisation theorem is not
completely proven, but see recent work in~\cite{Ji:2004wu}.

The \citeauthor{Collins:1993kk} mechanism exploits intrinsic quark motion
inside the produced hadron $h$. Thus, assuming factorisation to be valid, the
cross-section difference is
\begin{align}
  E_h \, \frac{\D^3\sigma( \Vec{S}_T)}{\D^3\Vec{P}_h} -
  E_h \, \frac{\D^3\sigma(-\Vec{S}_T)}{\D^3\Vec{P}_h}
  & = -
  2 \, | \Vec{S}_T | \, \sum_{abc}
  \int \! \D{x}_a
  \int \! \D{x}_b
  \int \! \D^2\Vec\kappa_T \,
  \frac{1}{\pi z}
  \nonumber
\\
  & \null \times
  \DT{f}_a(x_a) \, f_b(x_b) \,
  \Delta_{TT} \hat\sigma(x_a, x_b, \Vec\kappa_T) \,
  \DT^0 D_{h/c}(z, \Vec\kappa_T^2),
\end{align}
where $\Delta_{TT}\hat\sigma$ is a partonic spin-transfer differential
cross-section difference (see, \eg, \cite{Barone:2001sp}). The
\citeauthor{Sivers:1990cc} effect relies on $T$-odd $k_T$-dependent partonic
distribution functions and predicts \acp{SSA} of the form
\begin{align}
  E_h \, \frac{\D^3\sigma( \Vec{S}_T)}{\D^3\Vec{P}_h} -
  E_h \, \frac{\D^3\sigma(-\Vec{S}_T)}{\D^3\Vec{P}_h}
  & =
  | \Vec{S}_T | \,
  \sum_{abc}
  \int \! \D{x}_a
  \int \! \D{x}_b
  \int \! \D^2\Vec{k}_T \,
  \frac{1}{\pi z}
  \nonumber
\\
  & \null \times
  \Delta_0^T f_a(x_a, \Vec{k}_T^2) \, f_b(x_b) \,
  \frac{\D\hat\sigma(x_a, x_b, \Vec{k}_T)}{\D\hat{t}} \, D_{h/c}(z),
\end{align}
where $\Delta_0^T{f}$ (related to $f_{1T}^\perp$) is a $T$-odd distribution.
Finally, the effect studied by \citet*{Boer:1999mm} gives rise to an asymmetry
involving another $T$-odd $k_T$-dependent distribution $\DT^0f$ (related to
$h_1^\perp$) and a partonic initial-spin correlation differential cross-section
$\Delta_{TT}\hat\sigma'$:
\begin{align}
  E_h \, \frac{\D^3\sigma( \Vec{S}_T)}{\D^3\Vec{P}_h} -
  E_h \, \frac{\D^3\sigma(-\Vec{S}_T)}{\D^3\Vec{P}_h}
  & = -
  2 | \Vec{S}_T | \,
  \sum_{abc}
  \int \! \D{x}_a
  \int \! \D{x}_b
  \int \! \D^2\Vec{k}'_T \,
  \frac{1}{\pi z}
  \nonumber
\\
  & \null \times
  \DT{f}_a(x_a) \, \DT^0 f_b(x_b, \Vec{k'}_T^2) \,
  \Delta_{TT} \hat\sigma'(x_a, x_b, \Vec{k}'_T) \, D_{h/c}(z).
\end{align}

It was shown in \cite{Efremov:1982sh} that non-vanishing \acp{SSA} can also be
obtained in \ac{PQCD} by resorting to the gluonic poles present in higher-twist
diagrams involving $qqg$ correlators. Such asymmetries were evaluated in
\cite{Qiu:1991pp.x}, where direct photon production was studied, and for single
hadron production later in \cite{Qiu:1998ia}. This program was extended to
cover chirally-odd contributions in~\cite{Kanazawa:2000hz.x}. In the general
case one has
\begin{align}
  \D\sigma &=
  \sum_{abc}
  \left\{ \vphantom{D_{h/c}^{(3)}}
    G_F^a(x_a, y_a) \otimes f_b(x_b) \otimes
    \D\hat\sigma \otimes D_{h/c}(z)
  \right.
  +
  \DT{f}_a(x_a) \otimes E_F^b(x_b, y_b) \otimes
  \D\hat\sigma' \otimes D_{h/c}(z)
  \nonumber
\\[-1ex]
  & \hspace{14em} \null
  +
  \left.
    \DT{f}_a(x_a) \otimes f_b(x_b) \otimes
    \D\hat\sigma'' \otimes D_{h/c}^{(3)}(z)
  \right\}.
\end{align}
The first term does \emph{not} contain transversity and is the chirally-even
mechanism studied in \cite{Qiu:1998ia}; the second \emph{does} contain
transversity and is the chirally-odd contribution analysed in
\cite{Kanazawa:2000hz.x}; and the third contains transversity and a twist-3
fragmentation function $D_{h/c}^{(3)}$.

\begin{figure}[hbt]
  \centering
  \includegraphics[width=0.35\textwidth]{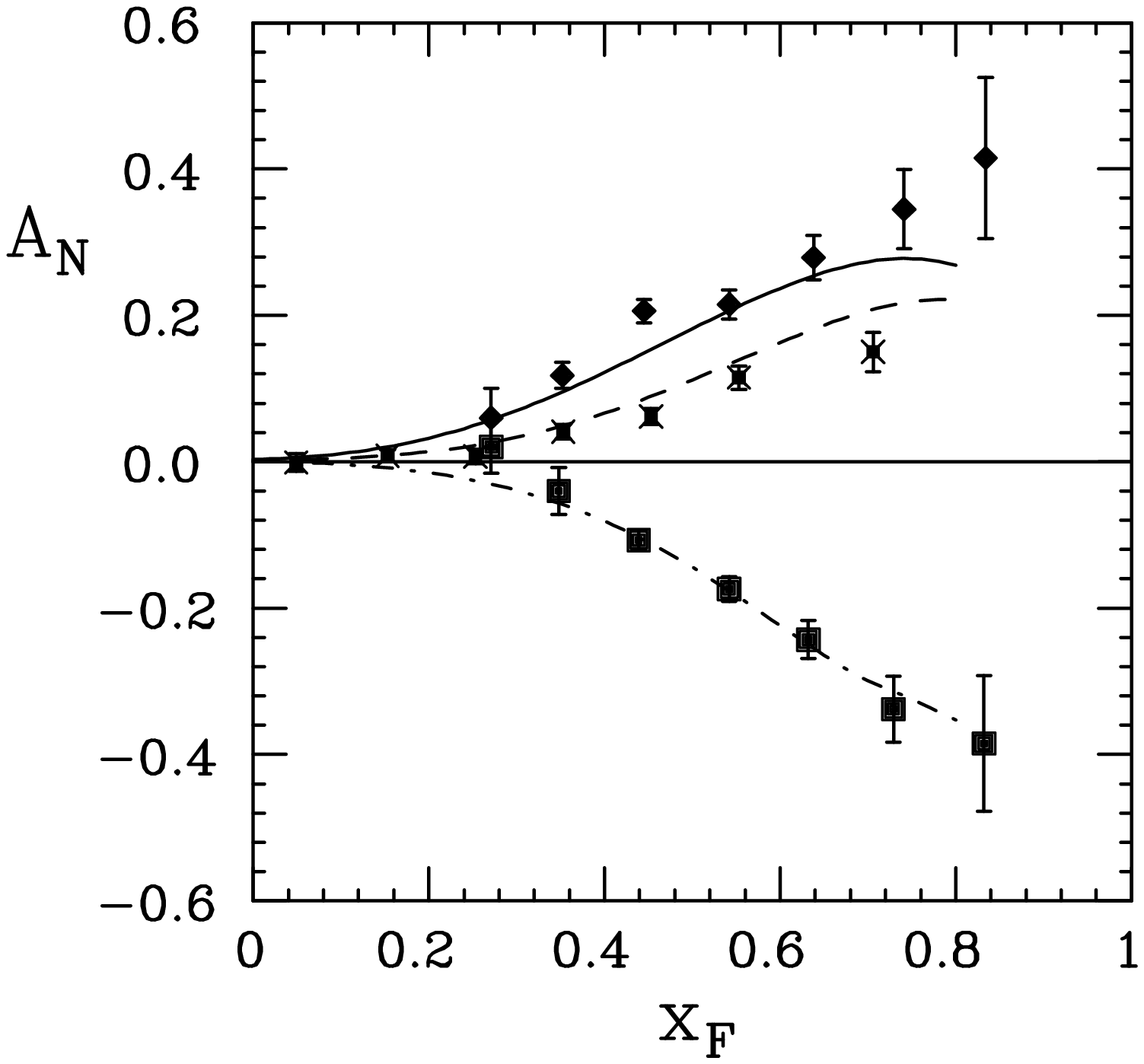}
  \qquad
  \raisebox{48.5mm}[0pt][0pt]{%
    \includegraphics[angle=-90,width=0.35\textwidth]{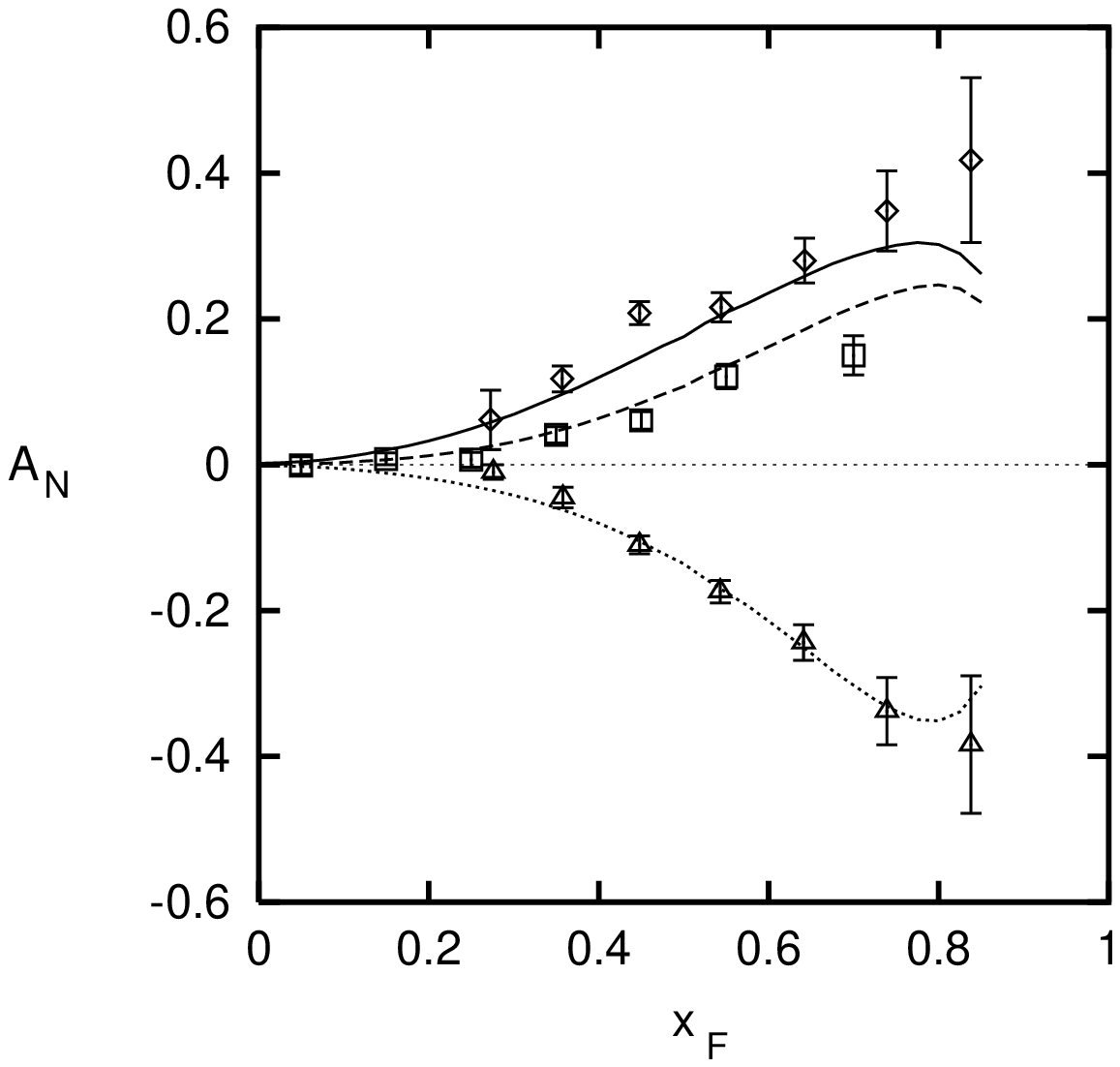}
  }
  \raisebox{6ex}{\makebox[0pt][r]{(a) \hspace{13em} (b) \hspace{7em}}}
  \caption{%
    Single-spin asymmetries
    Assuming (a) only \citeauthor{Sivers:1990cc} \cite{Anselmino:1998yz} and
    (b) only \citeauthor{Collins:1993kk} \cite{Anselmino:1999pw} mechanisms.
  }
  \label{fig:phenom}
\end{figure}
Models inspired by the previous possible ($k_T$-dependent) contributions
compare well to data in \cite{Anselmino:1998yz, Anselmino:1999pw}. Indeed,
phenomenological fits based on either the \citeauthor{Sivers:1990cc} or
\citeauthor{Collins:1993kk} mechanisms work equally well and are thus
impossible to distinguish at present, see Fig.~\ref{fig:phenom}. The
calculations based on three-parton correlators \cite{Qiu:1991pp.x} are opaque,
involving many diagrams, complex momentum flow, colour and spin structure. One
sees that unfortunately the current knowledge at both the theoretical and
experimental level do not permit a clear and concise description of these
phenomena.

However, the twist-3 correlators obey constraining relations with
$k_T$-dependent densities and they also exhibit a novel factorisation property.
It is this simplification that I shall now describe in more detail. The twist-3
diagrams involving 3-parton correlators supply an imaginary part via a pole
term \cite{Efremov:1985ip} (spin-flip is implicit).
\begin{wrapfigure}{r}{0pt}
  \makeatletter
    \let\@makecaption\AIP@makefigurecaption
  \makeatother
  \quad
  \begin{minipage}{3cm}
  \centering
  \includegraphics[width=\textwidth,bb=391 558 495 701,clip]
                  {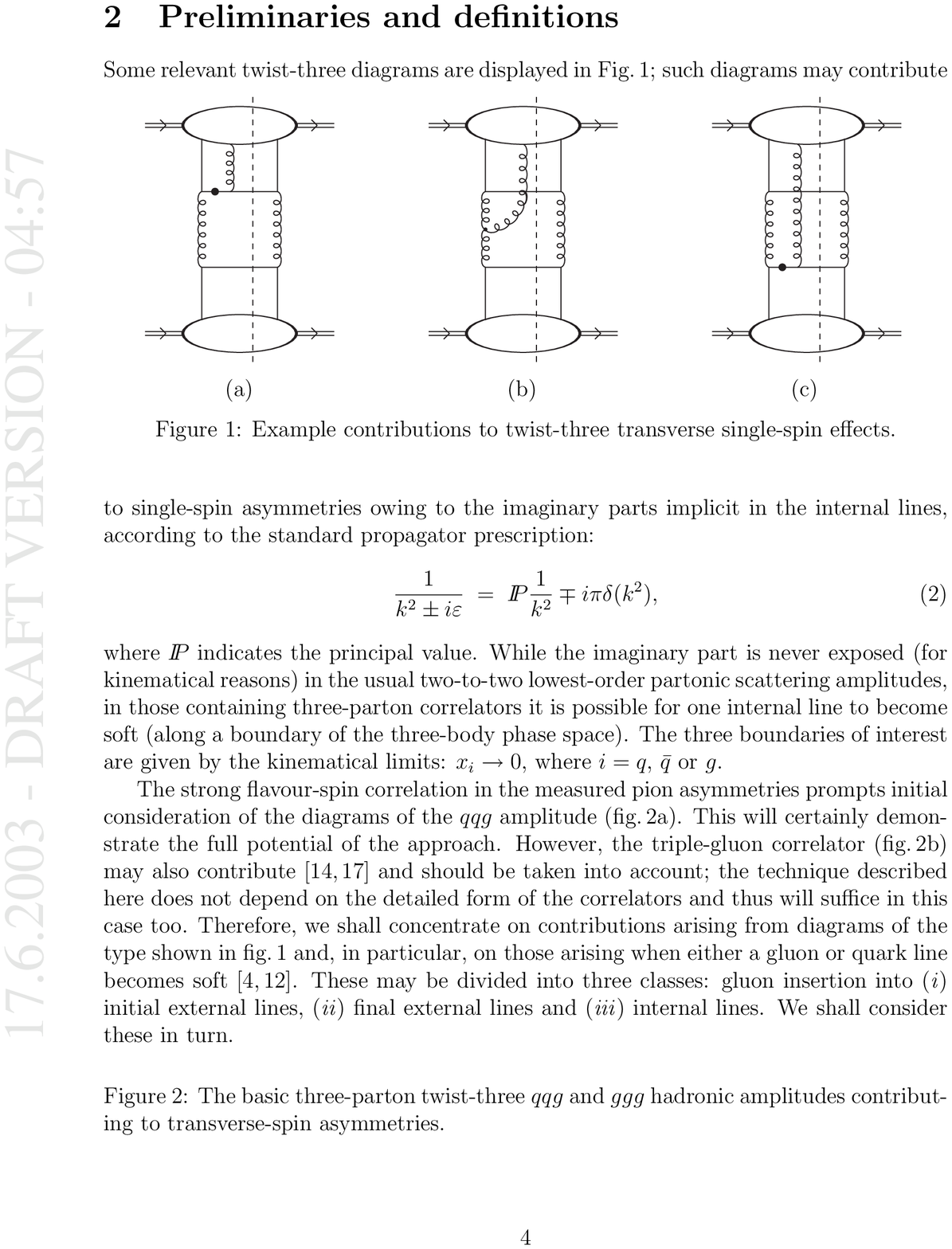}
  \caption{%
    Example of a dominant higher-twist diagram in the large-$\Nc$ limit.
  }
  \label{fig:dominant}
  \vspace*{-1.3ex}
  \end{minipage}
\end{wrapfigure}
The $\I\varepsilon$ propagator prescription (for
\,\rlap{$-$}$\bullet$\llap{$-$}\, in Fig.~\ref{fig:dominant}), leads to an
imaginary contribution for $k^2\to0$:
\begin{align}
  \frac{1}{k^2\pm\I\varepsilon} =
  \mathrm{I\!P} \frac1{k^2} \mp \I\pi\delta(k^2),
\end{align}
For a gluon, with momentum $x_gp$, inserted into an initial or final external
line $p'$, one has $k=p'-x_gp$, and thus $x_g\to0$. This can be shown
systematically for all poles (gluon and fermion), \ie, on all external legs
with all possible insertions~\citep{Ratcliffe:1998pq}.

This still generates very complex structures; there are many possible
insertions, with contributions of different sign and momentum dependence. The
colour structure of the various diagrams is also very different. In all cases
examined just one diagram dominates in the large-$\Nc$ limit (see
Fig.~\ref{fig:dominant}), \ie, other insertions into external legs are
suppressed by\\[-2.5ex]
\begin{wrapfigure}{l}{0pt}
  \makeatletter
    \let\@makecaption\AIP@makefigurecaption
  \makeatother
  \begin{minipage}{8.0cm}
  \centering
  \includegraphics[height=2.2cm,bb=142 661 250 749,clip]{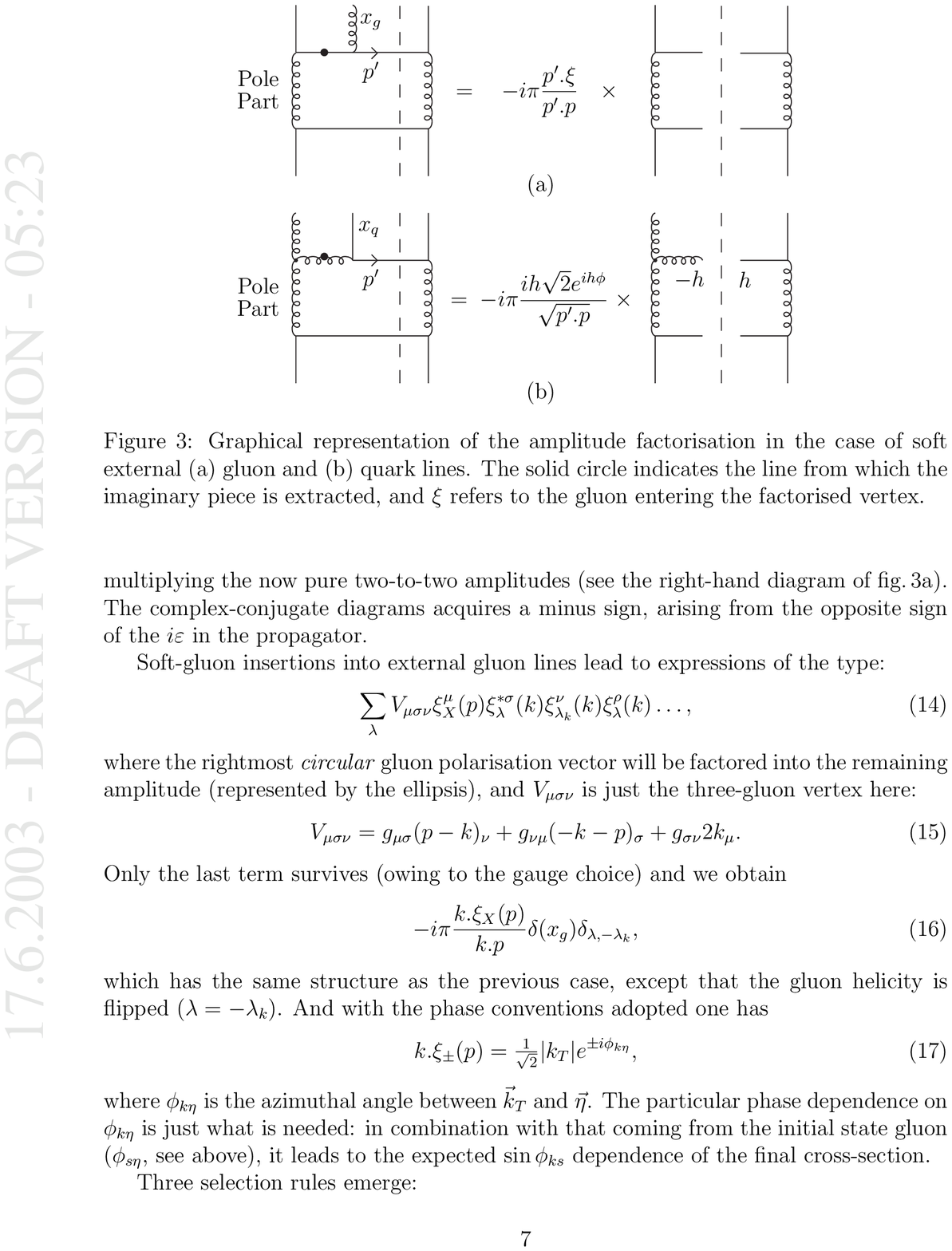}
  \includegraphics[height=2.2cm,bb=250 661 437 749,clip]{epsfiles/pole-factor}
  \caption{%
    A graphical representation of the higher-twist pole-factorisation
    mechanism.
  }
  \end{minipage}
  \hspace*{1em}
\end{wrapfigure}

\noindent%
$1/\Nc^2$. Assuming that similar simplifications can also be found for the
$k_T$-dependence, then by relating the higher-twist to the $k_T$-dependent
mechanisms via the equations of motion, unique predictions may be possible for
azimuthal \acp{SSA}. A study along these lines has indeed already been made for
\ac{DY} in \cite{Ma:2003ut}; predictions were, however, found \emph{not} to be
unique there. The $k_T$-dependence and higher-twist connections have also been
exploited in~\cite{Bacchetta:2004zf}.

\section{Summary and Conclusions}

Transversity is now considered equally important an aspect of nucleon structure
as the other two leading-twist parton densities and a complete description of
the nucleon requires its understanding. Theoretically, all the pieces of the
\ac{PQCD} puzzle (up to \ac{NLO}) are in place and lattice calculations
indicate transversity to be sizable. On the experimental side, unfortunately,
there are as yet no real data. However, the future is promising and before long
we shall start to harvest interesting results. The phenomenology, while not
dissimilar to the other leading-twist structures, has interesting
peculiarities: evolution is non-singlet (so analysis should be cleaner),
spin-half objects can contain gluonic transversity but it is not accessible via
the usual partonic hard-scattering processes (this should perhaps be examined)
and there is no associated sum rule (thus \ac{QCD} evolution is faster). This
all suggests that transversity could, in principle, allow clean extraction of
$\alpha_s$ from the evolution fits to the scale violating $Q^2$ variation (\cf
unpolarised).

\Acp{SSA} have progressed from having essentially no (\acs{PQCD}) theory to
almost too much! Hopefully, the multitude of mechanisms can be reduced to just
a few simple terms: experiment can eliminate some possibilities if null results
are obtained; relationships between three-parton correlators and
$k_T$-dependent densities should show up equivalences between apparently
different phenomenological models; while pole-factorisation and the large-$\Nc$
limit can simplify calculations and allow a clear pattern to emerge.
\bibliography{pigrostr,pigrotmp,pigropgr,pigrodbf,pigroxrf}
\bibliographystyle{pigroxue}
\end{fmffile}
\end{document}